\documentclass[doublecol]{epl2} 

\newcommand{\beq}{\begin{equation}}
\newcommand{\eeq}{\end{equation}}
\newcommand{\beqa}{\begin{eqnarray}}
\newcommand{\eeqa}{\end{eqnarray}}
\usepackage{xcolor}

\title{Activated dynamics of the Ising $p$-spin disordered model with finite number of variables}
\shorttitle{Dynamics of the Ising $p$-spin disordered model} 

\author{Daniel A. Stariolo\inst{1,2} \and Leticia F. Cugliandolo\inst{1}}
\shortauthor{D. A. Stariolo \& L. F. Cugliandolo}

\institute{                    
  \inst{1} Sorbonne Universit\'e, Laboratoire de Physique Th\'eorique et Hautes Energies, 
  UMR 7589 CNRS, Tour 13, 5\`eme Etage, 4 Place Jussieu, F-75252 Paris 05, France
  \\
  \inst{2} Universidade Federal Fluminense, Departamento de F\'isica and 
  National  Institute of Science and Technology for Complex Systems, Av. Gal. Milton Tavares de Souza s/n, 
  Campus da Praia Vermelha, 24210-346 Niter\'oi, RJ, Brazil
}

\abstract{
We study the dynamic and metastable properties of the fully connected Ising $p$-spin 
model with finite number of spins, with a focus on activated dynamics and
trap-like characteristics.
We propose a definition of trapping regions based on purely dynamical criteria.
We compute trapping energies, trapping times and
self correlation functions 
and we analyse their statistical properties in comparison to the predictions of
the well-known Bouchaud trap model. }

\begin{document}

\maketitle

\section{Introduction}

Models with quenched random potentials  describe physical systems with 
frozen impurities that alter the properties of the host material.  However, their relevance goes well 
beyond this field as they are also toy models for combinatorial optimisation, 
ecological stability, or even social sciences.

In physical applications, focus is set upon the thermodynamic limit
in which the number of degrees of freedom, say $N$, diverges.
In other applications,
$N$ is finite and it is necessary to understand strong finite size effects.
 In the context of the glass transition, usual mean field approaches
  are not able to describe the low temperature dynamical regime because the size
  of free-energy barriers diverges with $N$, while in finite dimensions they do not and
  activation is possible. A possible strategy to get closer to the behaviour of real glassy systems, which
we will pursue here, is to consider finite size mean field models.
With this aim, we will study the Ising (Boolean variables)
disordered $p$-spin model~\cite{Derridaprl1980} with finite $N$. 
Among the standard quenched random potential models, 
this one occupies a very important position. On the one hand, it is the 
standard mean-field model for the glass transition~\cite{KiTh1987b} and, on the other hand, it is 
intimately related to the celebrated K-sat problem of combinatorial optimisation~\cite{Monasson1999}. 
Beyond these two, this model also appears in studies of relaxation in fitness
landscapes of interest in several biological systems,
among others~\cite{Weinberger1993}.
Concretely, we will investigate the dynamical properties of the Ising $p$-spin model with small number of 
degrees of freedom at low temperatures, where activation over barriers is the dominant mechanism for relaxation,
and we will analyse the results in the context of well-known trap models of activated relaxation.


The presentation is organized as follows.
We start with a short introduction of the $p$-spin model and some of its well-known 
properties, its phases and transitions, in the $N\to\infty$ limit.
Next, we recall the definition and properties of the trap models that mimic activated dynamics in disordered 
systems. We then proceed to the analysis of the low temperature relaxation of finite size $p$-spin model. 
 Our aim is to characterise the trapping configurations
 reached via activation, focusing on small size systems to access  the interesting time regimes 
 with moderate numerical effort.
The measurements will help to identify common features  and also differences with the relaxation 
 of exactly solvable trap models.
Finally, we will draw some conclusions and point out some possible routes to pursue
this line of research.

\section{The $p$-spin models}

The Ising spin model with multi-spin random interactions~\cite{Derridaprl1980,Gross1984,Gardner1985,KiTh1987b} is defined by the energy function
\begin{equation}
\label{phamilton}
H_J[\{S_i\}] = -\!\!\!\!\!\!
\sum_{i_1<\dots< i_p = 1}^{N} J_{i_1, \dots, i_p} S_{i_1}\cdots S_{i_p} 
\; ,
\end{equation}
where $S_i=\pm 1$,  $i=1\ldots N$, 
and $J_{i_1, \dots, i_p}$ are independent identically distributed (i.i.d.)
quenched Gaussian random exchanges with zero mean and standard deviation $\sqrt{p!/(2N^{p-1})}$. 
The tensor of coupling
constants $J_{i_1, \dots, i_p}$ is symmetric under arbitrary permutations of the indices $\{i_1, i_2, \dots, i_p \}$
and it connects all possible groups of different $p$ spins. The model is therefore defined on a complete hyper-graph. A coupling to a heat bath is mimicked with single spin flip 
Monte Carlo (MC) dynamics that induce a nearest-neighbour random walk 
on the  $N$-dimensional hypercubic configuration space. 

\subsection{Infinite size behaviour}
The model has been much studied in its continuous version, in which 
a spherical constraint on real valued variables allows one to derive exact results for its
equilibrium thermodynamics~\cite{CrSo1992}, metastable properties~\cite{CrSo1995,PhysRevB.57.11251}
and non-equilibrium relaxation~\cite{PhysRevLett.71.173}. 
The Ising version was also considered in quite some detail and we 
summarise below the main features found so far. 

The {\it equilibrium} properties can be derived exactly in the $p \to \infty$ limit~\cite{Gross1984,Gardner1985}
with  microcanonical and canonical methods complemented by the replica trick, and are found to be 
equivalent to the ones of the Random Energy Model (REM)~\cite{Derridaprl1980}. 
For any $p\geq 3$ there is a static transition at a
temperature $T_s$ between a replica symmetric (RS) high temperature paramagnetic phase and a one-step replica symmetry breaking (1RSB) 
low-temperature glassy phase~\cite{KiTh1987b,Stariolo1990,Talagrand2003,Nakajima2008,Agliari2012a,Janis2015}. The transition is discontinuous, with a jump in the Edwards-Anderson order parameter, 
but no latent heat. Perturbative approximations for $p=2+\epsilon$~\cite{Gardner1985} showed that below the Gardner 
temperature $T_G<T_s$ the 1RSB solution is replaced by a full replica symmetry breaking (FRSB) one. 
Both $T_G$ and $T_s$ depend on $p$. The equilibrium energy density is 
plotted as a function of $T$ with a dashed-dotted (blue) line in Fig.~\ref{fig:thresholds}.

The {\it free-energy landscape} is rugged and complex.
Below a temperature $T_d \, (>T_s)$ the Gibbs measure decomposes in 
an exponentially large  in $N$ number of  metastable states~\cite{Rieger1992,Oliveira1997}. 
These have free-energy densities
between $f_{\rm eq}$, the equilibrium one, and $f_{\rm max} > f_{\rm eq}$, and they can be further 
distinguished according to their stability~\cite{Montanari2003,Crisanti2005a,Rizzo2013a}. 
A careful analysis of the complexity~\cite{Monasson1995}
suggested that at $T\in [T_G,T_s]$ 
the metastable states are of two kinds:
above a limiting value, $f_G$, they are marginal (in technical terms, FRSB would be needed to calculate the complexity)
while the ones below $f_G$ are not (1RSB is fine)~\cite{Montanari2003,Crisanti2005a,Rizzo2013a}. 
In more intuitive terms, metastable states 
are not properly separated minima above $f_G$ and, as 
$f_G=f_{\rm eq}$ at $T_G$, below this temperature all metastable states are connected through flat directions.
At $T_s$, $f_G=f_{\rm max}$. 

In the $N\to\infty$ spherical model, the {\it relaxation dynamics}  
from an equilibrium initial condition  at any $T>T_d$
to any $ T<T_d$ occurs out of equilibrium and 
approaches a flat threshold level in the free-energy landscape~\cite{PhysRevLett.71.173}. 
Accordingly, the  energy density decays algebraically
towards a threshold energy $e_{\rm th}$. 
For longer time scales, expected to scale exponentially with $N$, activation over free-energy barriers 
should let the energy density go below $e_{\rm th}$ in a much slower way.
The heuristic image is one in which the system performs a sequence of jumps between different valleys at random times, 
with rates governed by the heights of connecting passes or saddle points. 
For the moment, the full analytic treatment of this regime remains out of reach. 
Evidence for trapping  is provided by 
$N\to\infty$ calculations in which the system is initially prepared in a sub-threshold metastable state 
and it is seen to remain in it ever after~\cite{Barrat1996b,Franz1995a}.

In the thermodynamic limit, the Ising p-spin model also undergoes a dynamic transition at $T_d>T_s$~\cite{KiTh1987a,KiTh1987b,Ferrari2012}
to an out of equilibrium phase 
that could only be studied analytically
with soft spins and Langevin dynamics.
A similar phenomenology to the one of the spherical case is expected, with additional complications 
due to the existence of marginally stable states below the threshold.  The 1RSB threshold
energy density is 
plotted as a function of $T$ with a continuous (blue) line in Fig.~\ref{fig:thresholds}.

\subsection{Finite size behaviour}

The  relaxation of mean-field disordered systems 
with free-energy landscapes plagued with metastable states is 
expected to be driven by activation in time scales  scaling exponentially with the system size.
We are not aware of numerical simulations that study the interplay between
finite $N$ and finite times in the out of equilibrium properties of the $p$-spin Ising model. 
However, the relaxation of the closely related random orthogonal model for finite $N$
was shown to undergo a cross-over from smooth to 
activated behaviour~\cite{Crisanti2000,Crisanti2000a}. The problem has gained renewed attention
recently, specially after the appearance of some rigorous results showing that the REM model
~\cite{Derridaprl1980} behaves as a trap model~\cite{Bouchaud1992} upon a coarse graining of
the time scales of observation~\cite{Arous2002,Arous2008,Gay2016,Cerny2017}.


\subsection{Traps models (TM)}

  These are a family of toy models~\cite{Dyre1987,Bouchaud1992,BouchaudDean1995,Monthus_1996}
  in which a coherent
subregion of the system is schematically described by the motion
of a point evolving in a landscape of traps, separated by barriers that can 
only be overcome via activation. The wandering of this point is described
by a master equation with hopping rates
that encode the statistics of valley depths, barrier heights and the geometry
of phase space. 

In their simplest realisation, M traps have i.i.d.
energies chosen 
from an exponential probability distribution function (pdf)~\cite{Dyre1987,Bouchaud1992,BouchaudDean1995}
  \beq
P(E_i) = \lambda \ \exp{[\lambda\,(E_i-E_0)]} \; , \hspace{1cm} E_i \leq E_0
\; ,
\label{eq:exp-dist}
\eeq
with $\lambda$ a parameter and $E_{0}$ a fixed escape energy level
(usually, $E_{0}=0$). An Arrhenius argument for the mean time
to leave the $i$th trap yields:
\beq
\tau_i = \tau_0 \ \exp{[\beta \,(E_{0}-E_i)]}
\; , 
\eeq
where $\beta = 1/T$ is the inverse temperature. 
Once at $E_{0}$ 
the process starts anew and all traps are equally likely to be accessed: the history of past events
is totally forgotten. These
features lead to an algebraic distribution of trapping times, 
\beq \label{eq:trap-times}
\phi(\tau;x) 
\propto \tau^{-(1+x)}
\; ,
\eeq
where $x=\lambda /\beta=T/T_c$. For $T<T_c$ one time averages diverge at long times, the 
system never reaches equilibrium and the dynamics shows ageing.
An important generalization of the model, which takes it closer to model glass formers
and disordered spin systems involves
Gaussian distributions of trap energies~\cite{Monthus_1996,Cammarota2018}.
A connection with the exponential TM and rigorous proofs of the realisation of 
TM dynamics in the REM (equivalent to the $p\to\infty$ case),
with  microscopic transition rates that depend only on the departing configurations have been
given in~\cite{BenArous2003,Bovier2003}. Extensions to  finite $p$, where energies are
correlated,
though still within 
the simplified microscopic dynamics, were later presented in~\cite{Arous2008a}.

Some defining characteristics of the TM are strongly simplified with respect
to more realistic models: (i) trap energies are i.i.d. random
variables, chosen from a predefined pdf; (ii) each
trap can be reached from any other in a single jump,
in this sense, the landscape is completely
connected, and recurrent trap visits do not take place; 
(iii) the system has to jump to a
 fixed energy level $E_{0}$ to leave a trap, in other words, the transition rates depend only on the 
 energy of the departing trap.

 An interpolating rule for the transition rates that lifts the last restriction and takes
 into account the 
 energy of the arrival trap was proposed in~\cite{Rinn2000}  and recently
 studied in~\cite{Cammarota2018}.
Recently, rigorous results for the
Metropolis dynamics of the REM were also derived~\cite{Gay2016,Cerny2017}. 

Special challenges have been found
while trying to confront the theoretical results with computer simulations. An important piece of
information is that, in order to observe TM-like behaviour in models with Gaussian energies,
the dynamics have to be
``coarse-grained'', meaning that relevant traps do not correspond to single configurations but
to a bunch of them, reminiscent of the ``metabasins'' scenario used to describe the relaxational
dynamics of supercooled liquids~\cite{Doliwa2005}. This is equivalent to a renormalisation of
the time scales of observation in order to recover independence of individual jump events between
traps,
i.e. the {\em renewal} property of the Markov process, a critical assumption in the theoretical works
~\cite{Arous2002,Arous2008}. Interpreting the outcome of simulations of
the REM and generalisations has proved to be a very complex task. Convincing evidence for
an effective TM description emerges only when the observation times are scaled
exponentially with system size 
$N$~\cite{Baity-Jesi2018,Baity-Jesi2018a,Cammarota2018}.

\section{Results}

We fixed $p=3$ in (\ref{phamilton}), with system sizes $20 \leq N \leq 200$ and temperatures
$0.1 \leq T \leq 0.5$.
The static and dynamic critical temperatures in the large $N$ limit are 
$T_s\approx 0.651$, and $T_d\approx 0.682$.
Typical disorder averages were taken over 
1.5 $10^5$ coupling realisations (denoted $[\dots ]$). We used single spin flip
Metropolis dynamics and the unit of time is the Monte Carlo step (MCs),
a step being $N$ attempts to flip randomly chosen spins.
  
 \subsection{Threshold, activation and equilibrium}
 
\begin{figure}[t!]
\centering
\includegraphics[scale=0.69]{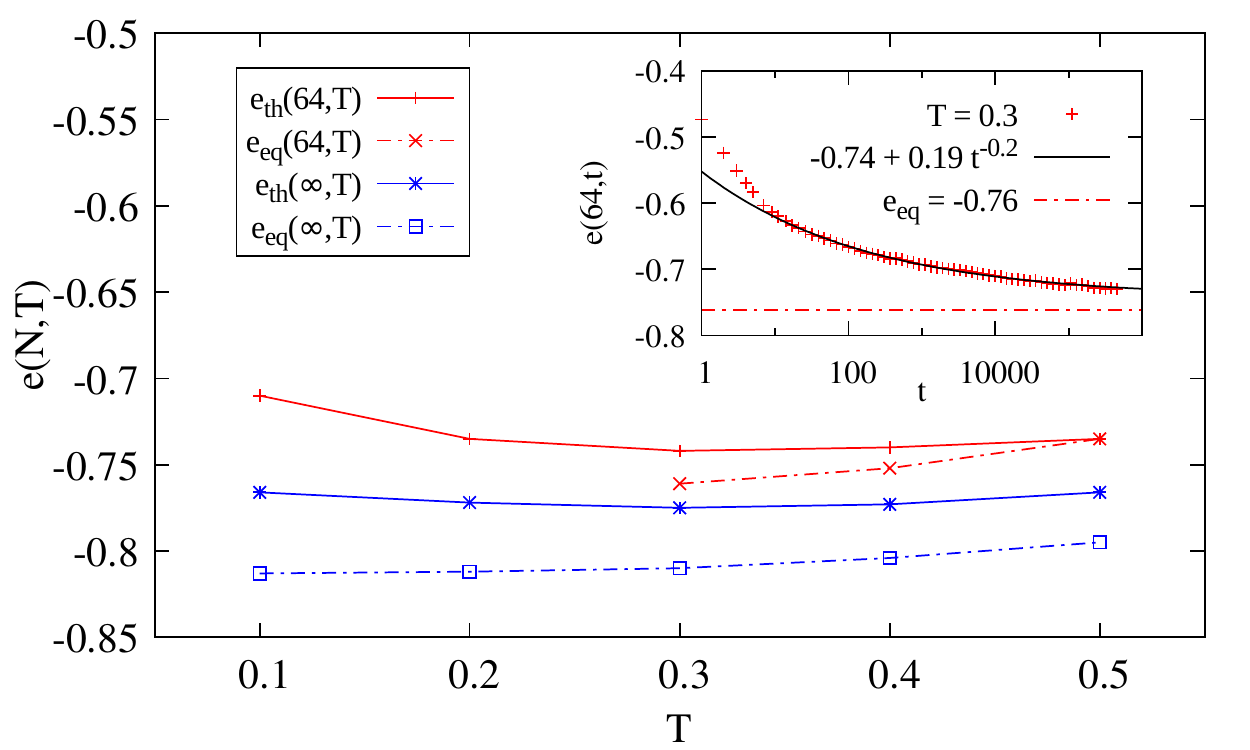}
\caption{(Colour online.) Threshold and equilibrium energy densities as functions of temperature for
  $N=64$ (simulations) and $N=\infty$ (analytic), as specified in the legend. In the inset, an
  example of the power law relaxation to the threshold level in the $N=64$ system at $T=0.3$.
The equilibrium data for $N=64$ are from~\cite{Billoire2005}. 
}
\label{fig:thresholds}
\end{figure}

We first monitored the relaxation of the disordered averaged internal energy density 
$e(t;T,N) \equiv [H_J(t)]/N$ after a quench from a completely disordered initial state.
For $N=64$ at the highest temperature, $T=0.5$, 
the dynamics reaches equilibrium  in time scales of the
  order of $10^6$ MCs, but it is unable to do it at lower temperatures within
  these time scales, as shown in Fig.~\ref{fig:thresholds}. Instead, the relaxation curve follows a power law decay, as shown in the inset
  for $T=0.3$. 
The asymptotic limit 
$
e_{\rm th}(T, N) \equiv \lim_{t\to\infty} e(t;T, N) 
$,
estimated from a fit of the finite-time data,
$
e(t; T, N) \simeq e_{\rm th}(T, N) + a \, t^{- b}
$,
gives us an empirical measurement of the energy density of the finite-$N$ threshold, approached 
following smooth regions of the free energy landscape. 
 In the main panel we show $e_{\rm th}(T,64)$ thus obtained at several low $T$'s together with
  the equilibrium values $e_{\rm eq}(T,64)$ reported in Ref.~\cite{Billoire2005} for $T \geq 0.3$. 
   In order to compare the finite $N$ energy scales with the diverging $N$ ones, we also plot 
  the analytical threshold (computed within a replica formalism at 1RSB) and the equilibrium energies of the $N \to \infty$
  system~\cite{Montanari2003}.
 
  Figure~\ref{fig:thresholds} shows that, for $N=64$, 
  the distance between the threshold and equilibrium energies increases as $T$ is 
  lowered (the same is found in the 1RSB replica calculation of~\cite{Montanari2003}, although
    that solution turns out to be unstable against further replica symmetry breaking).
In order to avoid
    getting stuck at the threshold level, we chose to work with  $N=20$, a choice that allowed us to extract interesting results even at very low
    temperatures $T \geq 0.1$.

 \subsection{Sub-threshold relaxation and evidence of trap behaviour}
 
\begin{figure}[t!]
\begin{center}
\includegraphics[scale=0.35]{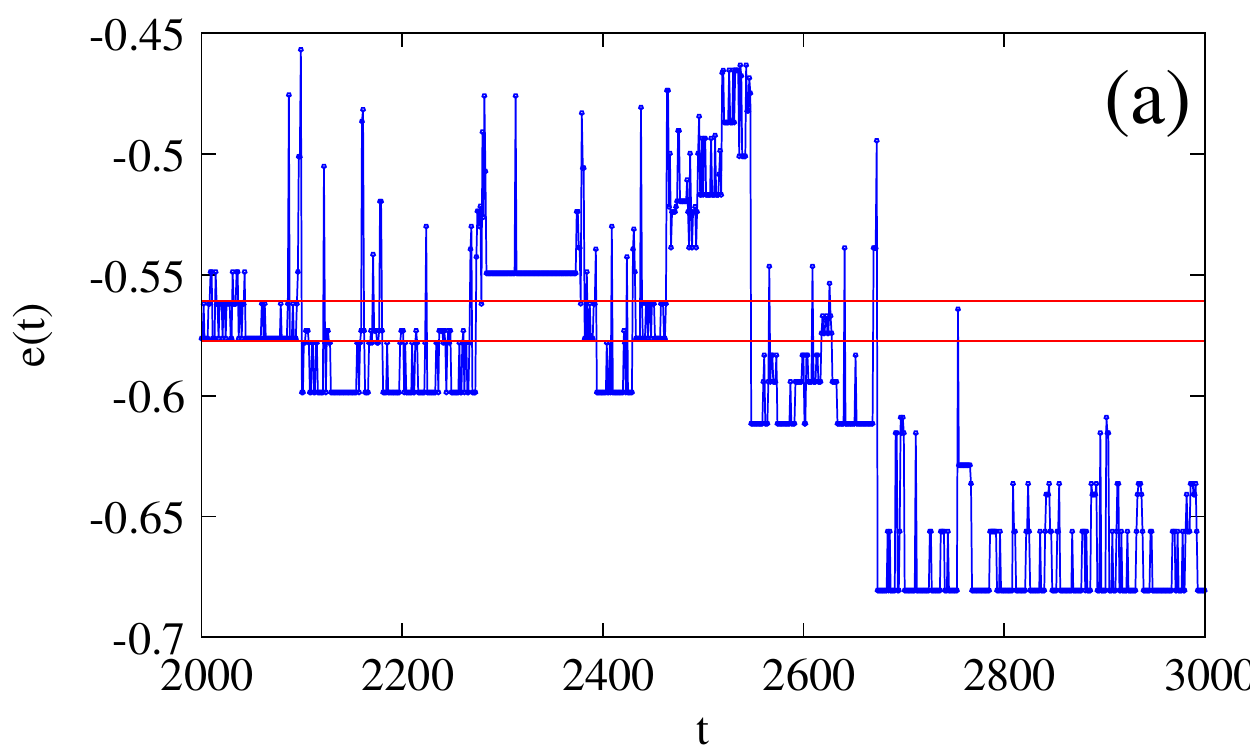}
\hspace{-0.4cm}
\includegraphics[scale=0.35]{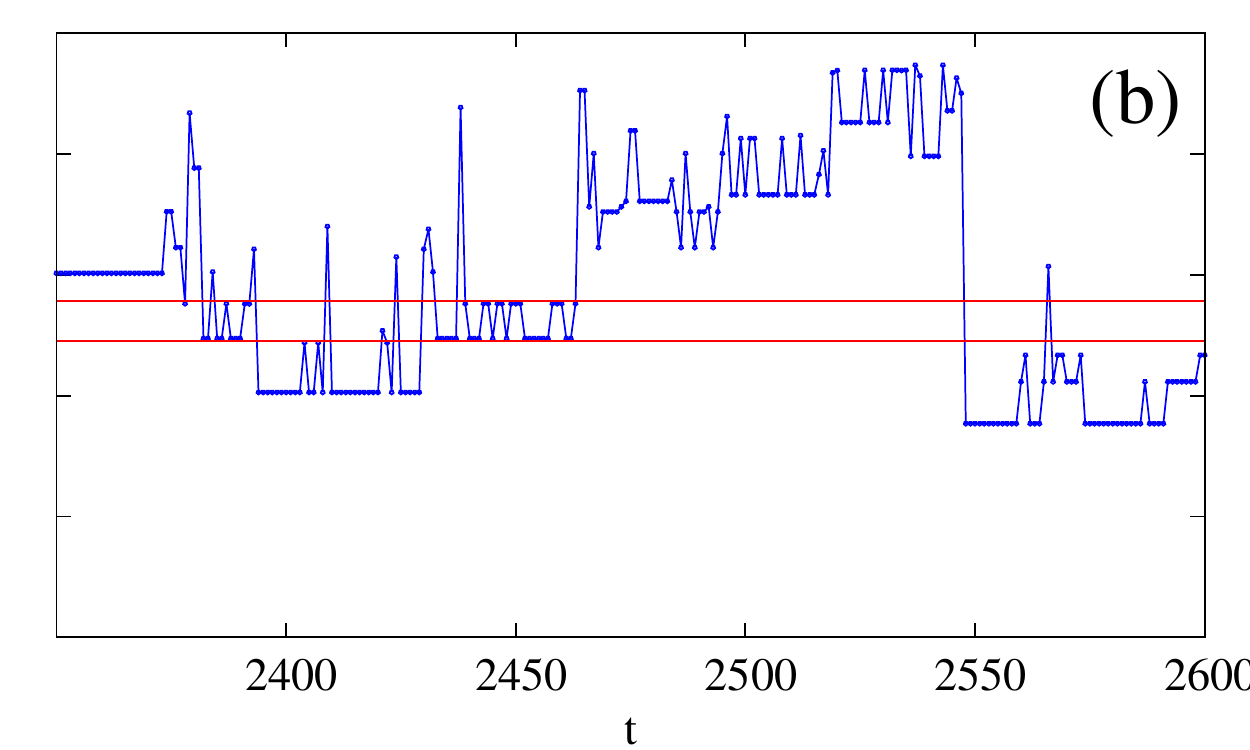}
\end{center}
\vspace{-0.35cm}
\caption{(Colour online.) (a) The energy density vs. time
  of a single run for a system with $N=20$ and $T=0.3$,
  over a time interval of length $10^3$ MCs. (b)
 A zoom over a restricted interval. A point per MCs is plotted.
 The two horizontal lines correspond to solutions of the corresponding TAP
 equations.
}
\label{fig:single-runs}
\end{figure}

In Fig.~\ref{fig:single-runs} we show the evolution of the energy density
of a single system after a quench to $T=0.3$ from a random initial condition.
The time scale of observation is considerably shorter than $t_{\rm eq}$. The microscopic
evolution shows
 the existence of trapping regions in
 configuration space.  As already observed 
in~\cite{Cugliandolo1997,Iguain2001,Berthier2003} (though with energy injection dynamics), 
the single trajectories show a self-similar pattern 
with  trapping and release motion.
On the one hand, this image demonstrates that  
there are recurrent visits to a set of configurations with some degree of dynamical stability,
differently from what happens in the basic TM. On the other hand, 
on the long run the system relaxes to lower energies in its way to equilibrium. It is also
  possible to see that 
the wandering in configuration space
proceeds essentially  {\em via activation events}.
This mechanism for relaxation, although
assumed to be important below the dynamic transition temperature $T_d$, has been hardly observed
 in simulations of realistic glassy models or even in experimental time scales. 
 
 The two horizontal lines in Fig.~\ref{fig:single-runs} are energy densities of 
 Thouless-Anderson-Palmer (TAP) states determined by
 an iterative solution~\cite{Bolthausen2014} of the TAP mean-field equations~\cite{Rieger1992}. Although for this 
 very small size subsequent correction terms to the Onsager one would have been expected to be needed to capture the 
 fixed points, we get energy levels that are in rather good agreement with the dynamic trapping
 energies with only the first correction.
 
 \subsection{Trap modelling}
 
From the above remarks it is tempting to try
to relate the relaxation of the Ising $p$-spin model with single spin flip Metropolis dynamics to the
one of TMs.
However, a major problem to define traps
 in the $p$-spin model is the lack of a fixed reference energy, $E_{0}$, which
 in the simple TM and its generalisations
 is a consequence of the static independence
of the energy levels~\cite{Baity-Jesi2018,Cammarota2018}.
Contrarily, in the finite $p$ model any two levels have
correlations that depend on the overlap between the corresponding spin configurations~\cite{Derridaprl1980}. 

\begin{figure}[t!]
\centering
\includegraphics[scale=0.55]{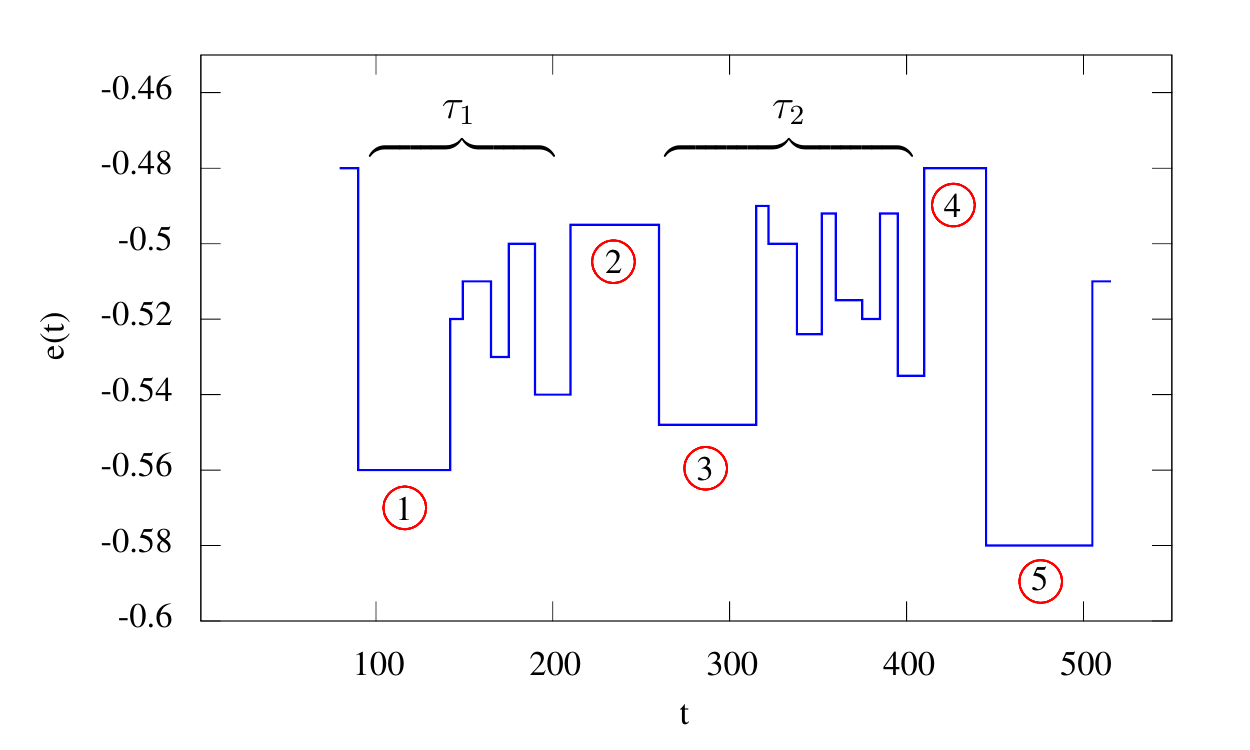}
\vspace{-0.2cm}
\caption{Schematic representation of the trap-like dynamics. }
\label{fig:schematic}
\end{figure}

Therefore, instead of defining traps relative to a fixed energy level, we adopted a
  definition based on dynamical stability that, roughly speaking,
  implements a time coarse graining and thus allows us to identify
  ``trapping regions'' in configuration space.  
  Let us consider a single MC run as the ones shown in 
  Fig.~\ref{fig:single-runs} and schematically represented 
 in  Fig.~\ref{fig:schematic} after filtering out fast fluctuations.
 A trap is defined as a sequence of configurations separated by two of them with a
 predefined
 ``large degree of dynamical stability''. We now give an operational definition of this concept
 with the help of two parameters, $\delta t_{\rm stab}$ and
 $\tau_{\rm min}$, measured in MCs. A configuration has a large degree of dynamical stability whenever it persists during
 at least $\delta t_{\rm stab}$. Configurations $1,2,\ldots 5$ in Fig.~\ref{fig:schematic} satisfy this
 condition for the chosen value of $\delta t_{\rm stab}$. Once one such configuration is found, we record the time $t_i$
 at which it was initially reached. We then follow the run until a different configuration 
 that also persists during at least $\delta t_{\rm stab}$ is reached, and we identify the time $t_f$ 
 when it first appeared after $t_i$. The time span
 between the appearance of one and the other configurations is $t_f-t_i$. If $t_f-t_i \geq \tau_{\rm min}$ 
 we identify the region in between them as a
 dynamical trap, with trapping time $\tau=t_f-t_i$ and energy $E_i$ (see Fig.~\ref{fig:schematic}).
 We put this construction to the test for different $\delta t_{\rm stab}$ and $\tau_{\rm min}$, and we recorded the
distribution of trap energies and times for each case. $\tau_{\rm min}$ fixes the minimum
time span of a trapping region, while $\delta t_{\rm stab}$ is a measure of the
dynamical stability of a particular configuration. In order to probe trapping
regions, as opposite to trapping configurations, it is necessary that $\tau_{\rm min} > \delta t_{\rm stab}$. 
We found strong constraints on the variability of these parameters, at least for the
system under study. For the lowest temperatures studied, $T=0.1, 0.2$ and $0.3$, if $\tau_{\rm min} \geq 500$
the only recorded trap energies are single low energy levels, irrespective of $\delta t_{\rm stab}$. Similarly, the counting
fails for $\delta t_{\rm stab} > 50$ because single configurations with such degree of stability are too rare
for the working temperatures. In all cases, within such constraints, the distributions of trap energies
are robust but some differences arise in the distributions of trap times, to be discussed
 below. The results shown in the following are for $\tau_{\rm min}=100$ and $\delta t_{\rm stab}=20$.

\subsection{Relaxation of the averaged trap energy}

\begin{figure}[t!]
\begin{center}
\includegraphics[scale=0.68]{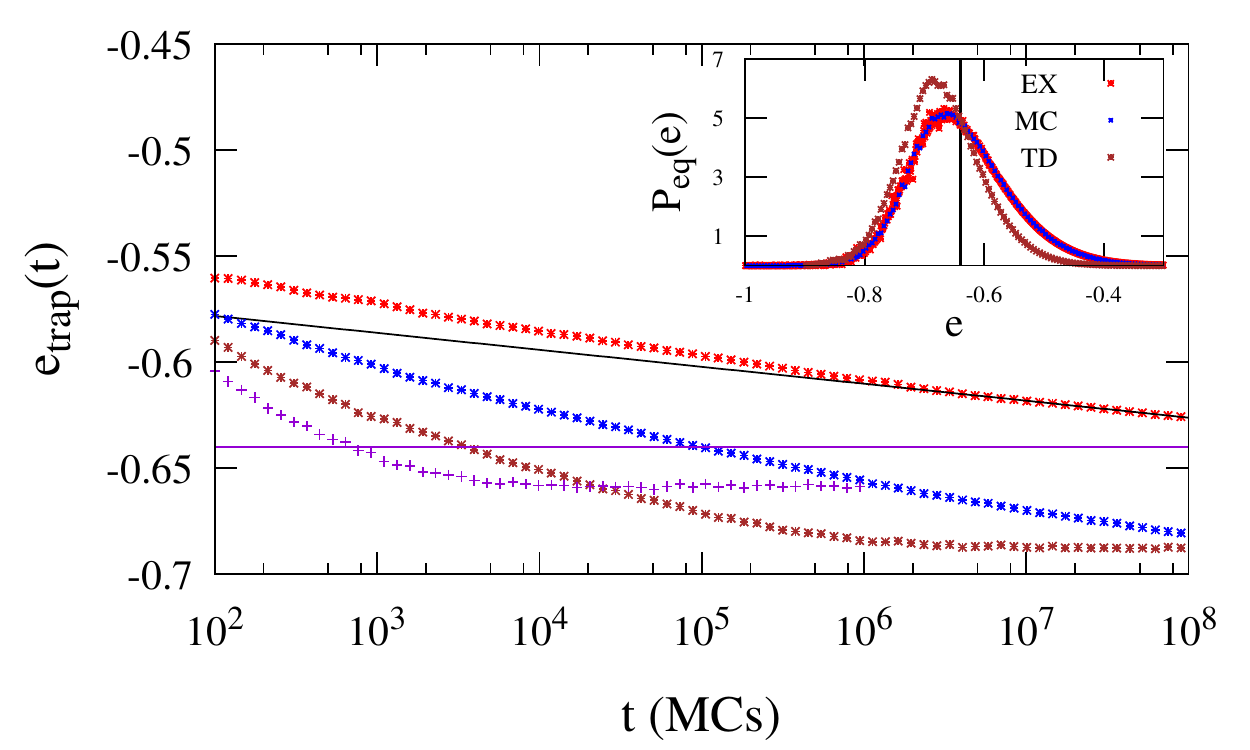}
\end{center}
\vspace{-0.3cm}
\caption{(Colour online.) The relaxation of the disorder averaged trap energy density of a system with $N=20$ spins
  at $T=0.1, \, 0.2, \, 0.3, \, 0.5$, from top to bottom. 
  The solid black line is a logarithmic fit to the late time data for $T=0.1$.
   The horizontal (purple) line is the equilibrium averaged energy density at $T=0.5$. 
Inset: The pdf of equilibrium energy densities. Data from exact enumeration 
and Monte Carlo simulations of $12 \times 10^4$ samples at $T=0.5$ shown with red and
blue datapoints (superposed).
The vertical black line is the averaged energy density.
The higher curve (in brown) is the pdf of trap energies in the stationary state. 
}
\label{fig:time-dep-energy}
\end{figure}

The main panel in Fig.~\ref{fig:time-dep-energy} (log-linear scale) shows
the relaxation of the average trap energy density, $e_{\rm trap}(t)$, 
following a quench to several low temperatures $0.1 \leq T \leq 0.5$. 
For the highest temperature, $T=0.5$, the evolution reaches equilibrium after 
$t_{\rm eq} \approx 10^4$~MCs. The equilibration time  at $T=0.3$ is
of the order of $5 \times 10^6$ MCs.
For the two lowest temperatures, $T=0.2$ and $T=0.1$,  
the relaxation is approximately logarithmic  at long times, $e(t) \simeq \Upsilon(T) \ln t/t_0$,
and the system is far from attaining equilibrium in the time scales of the simulation.
At $T=0.1$ the fit shown with a solid thin black line yields $\Upsilon(T) \approx 0.008$ which
is larger than the ideal
value $\Upsilon(T) = T/N = 0.005$ expected for a simple activation process.
Note that for this small system size a threshold energy level is absent. Indeed, its evaluation from an
algebraic fit to the early time relaxation gives unreliable small exponents as the dynamics
 soon crosses over to a logarithmic decay.

\subsection{The equilibrium energy density of $N=20$ systems}
 
For such small system sizes we can exactly enumerate all energy levels 
and obtain in this way the disorder dependent density of states $g_J(e)$. The product of this degeneracy and 
the Boltzmann factor 
${\rm e}^{-\beta Ne}/Z_J$, yields the equilibrium weight of the energy density $e$. In the inset of Fig.~\ref{fig:time-dep-energy} we compare this exact equilibrium 
pdf (red data) to the one sampled with MC dynamics beyond the equilibration 
time $t_{\rm eq}$ estimated from the relaxation of the energy density, at $T=0.5$ (blue data).
Having the energy pdf we can easily compute 
its equilibrium average, $\langle e \rangle_{\rm eq}(20,T)$, 
shown with a vertical solid line in the inset and with a horizontal
line in the main part of the
figure. Note that the equilibrium (mean) energy is slightly larger than the
most probable one. This is due to the asymmetric form of the pdf.
In the inset the
equilibrium pdf of the trapping energies, to be defined below, is also shown. 

\subsection{Distribution of trapping energies}

\begin{figure}[t!]
\centering
\includegraphics[scale=0.69]{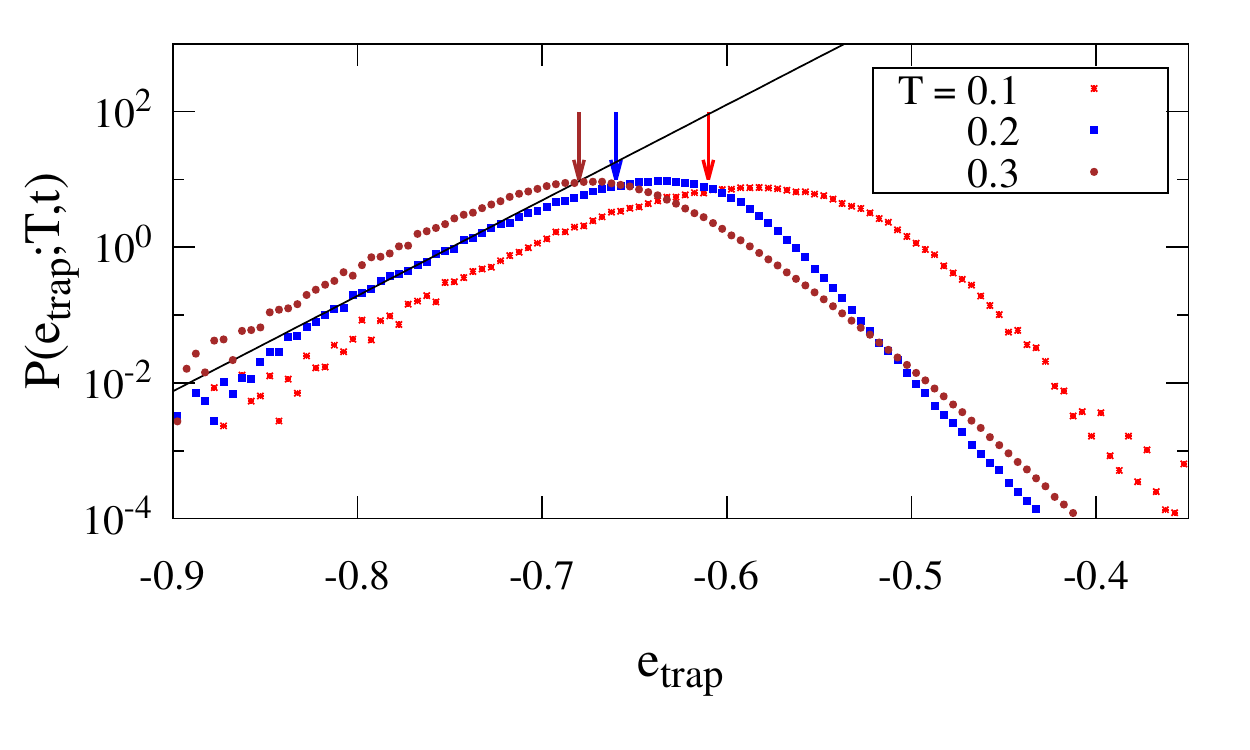}
\vspace{-0.7cm}
\caption{(Colour online.) The pdf's of trapping energy densities 
collected during $10^6$~MCs at each temperature.  The solid black line is an exponential fit to
  the left tail of the $T=0.2$ data. The arrows indicate the average
  energy densities in each case. 
    }
\label{fig:trap-energy-pdf}
\end{figure}

Figure~\ref{fig:trap-energy-pdf}
displays the pdfs of trapping energies, i.e. the energies of states like $1,3,\ldots $ in 
Fig.~\ref{fig:schematic} weighted by their trapping time, at $T=0.1, \ 0.2, \ 0.3$,  collected during $10^6$~MCs runs while the systems
are slowly relaxing to lower energies
 except for the  $T=0.3$ case which is approaching a stationary regime
  at this time scale. In spite of this difference, the three pdfs show a similar profile. 
  The small vertical arrows mark the average energy densities from Fig.~\ref{fig:time-dep-energy},
  which are a bit smaller than the most probable value except at $T=0.3$ where they almost
  coincide. 
  The most salient feature of these pdfs 
  is a low energy exponential tail that can be explained with an extreme value statistics argument.
  Indeed, the minima $E_{\rm min}$ of groups of $m$ i.i.d. Gaussian variables
  with zero mean and variance $\sigma^2$ follow a Gumbel distribution with an exponential tail
   with rate $\lambda=\sqrt{2\log{m}/\sigma^2}$ and typical energy
   $E_{\rm min}^{av}=-\sqrt{2\sigma^2 \log{m}}$~\cite{Bouchaud1997,Cammarota2018}. 
   For the $p$-spin model (\ref{phamilton}), 
    as the energies are (correlated) Gaussian
   random variables with zero mean and variance $\sigma^2=N/2$~\cite{Derridaprl1980},
   this argument would imply $\lambda = 2|e_{\rm min}^{av}|$. For general correlated
   random variables there are a few rigorous results regarding extreme
   value statistics~\cite{Carpentier2001,CluselBertin2008}, but no general framework yet. 
Although the values recorded with our procedure are not 
  necessarily minima over a large number of random independent energies, 
  we will check whether they conform to the Gumbel tail. 
  From exponential fits we obtain $\lambda \sim (1.25,1.50)$ and
  $e_{\rm min}^{av} \sim (-0.62,-0.72)$, which are in very good agreement with the extreme value statistics prediction.
  The pdfs also have a high energy tail which can be well fitted by another exponential, again with the exception
  of the data at $T=0.3$ for which the system is already near equilibrium. Both low and high energy 
  exponential dependencies have to be cut-off since the finite size energy density is bounded from below and above.

\subsection{Distribution of trapping times}

\begin{figure}[h!]
\centering
\includegraphics[scale=0.68]{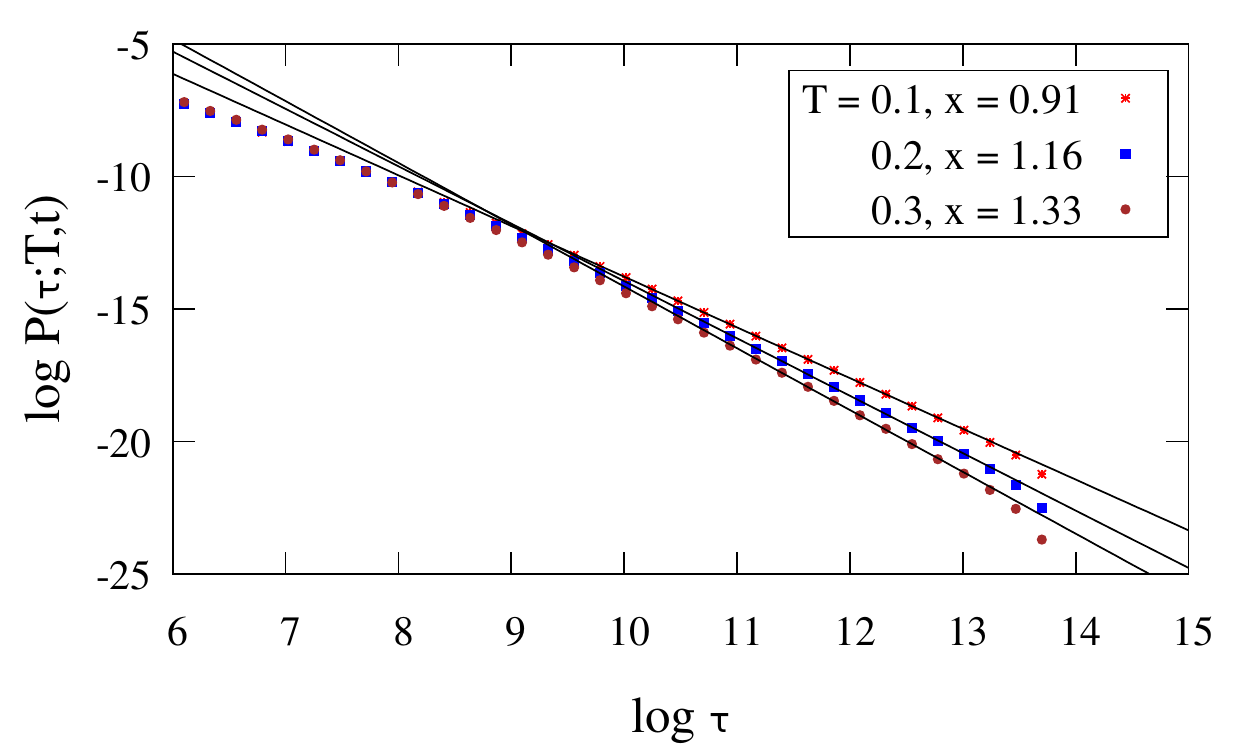}
\caption{(Colour online.) Double logarithmic representation of the trapping times pdfs 
  for the three temperatures shown in the legend. Solid lines are power law fits.
  The parameters $x(T)$ corresponding to the
standard power law form of the TM are shown in the legend.}
\label{fig:trap-times-pdf}
\end{figure}

We will attempt to compare our results to the (generalised) TM predictions, although
our trap definition is not immediately related to the one in the known TMs. 
  In Fig.~\ref{fig:trap-times-pdf} we show the distribution of trapping times for the same three
  $T$'s analysed before, together with algebraic fits to the long times sector.
  The first observation is that the pdfs are well represented by power laws at long trapping
  times, as in the TM.
The fitted values of the exponent $x$ in eq.~(\ref{eq:trap-times}) are $0.91 \leq x(T) \leq 1.33$,
and are given in the figure. 
  Changing parameters in the ranges $10 \leq \tau_{\rm min} \leq 500$ and
  $5 \leq \delta t_{\rm stab} \leq 50$, the exponents $x(T)$ vary as $0.89 \leq x(0.1) \leq 0.92$,
  $1.12 \leq x(0.2) \leq 1.15$ and $1.31 \leq x(0.3) \leq 1.34$. For $\tau_{\rm min} \geq 500$ or
  $\delta t_{\rm stab} \geq 50$ the process typically finds only one or two characteristic energies
   corresponding to the lowest and most stable states 
   ($\delta t_{\rm stab}\geq 50$) or, equivalently, the process identifies 
   only one or two very large traps ($\tau_{\rm min}\geq 500$),
  presumably the largest in a hierarchy of trapping regions.
  TM expectations correspond to $x$ smaller than one in the ageing
  regime, and this is consistently found at $T=0.1$. At $T=0.3$ the exponent is larger than 1 but the length of the 
  traps explored is also too close the equilibration time estimated from the average energy density relaxation.
  The case $T=0.2$ is borderline between these two. 
  Still, we insist upon the fact that there is no strong reason to expect a quantitative correspondence between
  the two models.


\subsection{Self correlation function}

\begin{figure}[b!]
\centering
\includegraphics[scale=0.68]{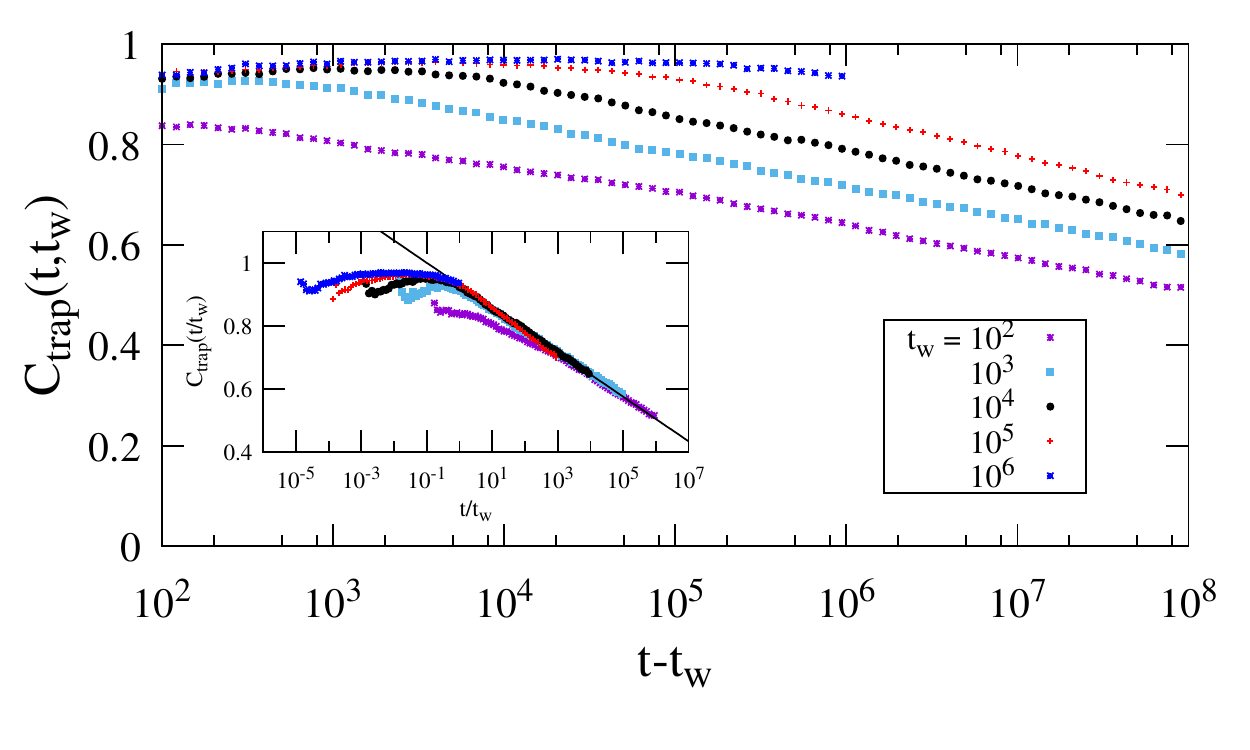}
\vspace{-0.5cm}
\caption{(Colour online.) The self correlation $C_{trap}(t,t_w)$ between trap configurations for different
  waiting times as described in the legend at $T=0.1$
  in log-linear scale. In the inset the same data as function of the scaling variable $w=t/t_w$. The
solid line is a logarithmic fit.}
\label{fig:trap-corr-pdf}
\end{figure}

The non-stationary ageing relaxation of disordered systems is usually characterised by the scaling properties of the
two-times self correlation $C(t,t_w) = N^{-1} \sum_{i=1}^N [\langle S_i(t) S_i(t_w) \rangle] $ 
(with the angular brackets indicating an average over initial conditions).
The most common scaling is the so-called ``simple ageing'' in which $C$ depends on $t$ and $t_w$ only through 
$t/t_w$, for $t$ and $t_w$ of the same order.
We defined a trap time-delayed self-correlation 
$C_{\rm trap}(t,t_w) = N^{-1} \sum_{i=1}^N [\langle S_i(t) S_i(t_w) \rangle] $.
At time $t_w$  we identified the last state $\{S_i(t_w)\}$ satisfying
($\delta t \geq \delta t_{stab}$).
We repeated this procedure at time 
$t$ and we computed the overlap between both states. Therefore, the configurations entering in the definition of trap
correlations are pairs of those used for computing the distributions of trap energies.
  The results obtained for $T=0.1$, a sufficiently low temperature so that
$t_{\rm eq} \gg t, t_w$, are displayed in Fig.~\ref{fig:trap-corr-pdf}. In the main panel a
linear-log plot of $C_{\rm trap}$ against time-delay for 
different waiting times is displayed, while in the inset the data are scaled as a function of $t/t_w$. 
The straight line is $\log(t/t_w)$ and 
fits the long time-delay data very well. In the context of the TM, an
  equivalent correlation function which measures the probability that the system did not leave a
  trap between $t_w$ and $t_w+t$ can be computed exactly and is known as the
  ``the Arcsin law'', $H_x(w)$~\cite{BouchaudDean1995,Cammarota2018}. Interestingly, the Arcsin law
  depends on the scaling variable $w=t/t_w$, the same which we found to describe quite well the $p$-spin model
  ageing behaviour. However, for large $w$, $H_x(w)$ decays as a power law
  $(t/t_w)^{-x}$, differently from the slower logarithmic decay that we
  observe in the small size $p$-spin model (inset in Fig.~\ref{fig:trap-corr-pdf}).

\section{Conclusions}

We analysed the relaxation dynamics of the Ising $p$-spin disordered model
in the context of the
well-known TM paradigm. We showed that for small sizes and at low temperatures it is
possible to identify a trap-like phenomenology whose most salient feature is the dominance of activated relaxation
events. Although a trap-like behaviour is evident at a qualitative
level, the very definition of a trap poses a great challenge in the $p$-spin model. At the center
of the difficulty lay the strong static correlations between energy levels, absent in the TMs. 
Another problem is that the time scales for relaxation in the interesting regime grow exponentially with system size. This
restricts the numerical studies to very small system sizes, which nevertheless show a highly non trivial phenomenology. Instead of the usual {\em energetic} definition of a trap, here we
proposed a {\em dynamical} one, which naturally implements a time coarse graining. 
We found trap energy pdfs with exponential
low energy tails, probably originated in an extreme value statistics
process induced by the working definition of traps. While this is interesting in itself, leading
to a resemblance with the exponential TM, the $p$-spin pdfs also show a high energy tail
falling-off slower than a Gumbel law, which would be the behaviour induced from an extreme value
process with i.i.d. Gaussian variables.
The  trapping time distributions have an algebraic 
decay, again in qualitative agreement
with TM predictions. The exponents vary between a value that is distinctively lower than one
at $T=0.1$ and one that is higher than one at $T=0.3$. We attribute the latter to the fact that the 
system is already exploring traps with life-time of the order of the equilibration time at this temperature.
A high energy sector in the pdf of trapping energies develops with time, with equal 
statistical weight as the low energy one. This may  be connected to the development of the threshold level, 
expected at larger system sizes, but practically absent in the small system studied in this
work. Finally, ageing correlations are also reminiscent of those of the
TM: at large time scales they depend on the same scaling variable $t/t_w$, although the
scaling function, which is essentially a logarithm in the $p$-spin model, is slower than the Arcsin
law characteristic of the TM. A more stringent comparison between the trap definition and
the free energy metastable states emerging from the TAP approximation may be an interesting route
to pursue further the relation between complex spin glass model dynamics and exactly solvable
models of the TM kind.

\acknowledgments
We are grateful to L. Dumoulin for very useful discussions at the early stages of this work. 
LFC is a member of Institut Universitaire de France. DAS acknowledges hospitality from LPTHE
where this work, financed in part by the Coordena\c c\~ao de Aperfei\c coamento de Pessoal de
N\'ivel Superior - Brasil (CAPES) - Finance Code 001, was done.


\begin{thebibliography}{10}
\expandafter\ifx\csname url\endcsname\relax\def\url#1{\texttt{#1}}\fi

\bibitem{Derridaprl1980}
\Name{Derrida B.} \REVIEW{Phys. Rev. Lett.}{45}{1980}{79}.

\bibitem{KiTh1987b}
\Name{Kirkpatrick T.~R. \and Thirumalai D.} \REVIEW{Phys. Rev.
  B}{36}{1987}{5388}.

\bibitem{Monasson1999}
\Name{Monasson R., Zecchina R., Kirkpatrick S., Selman B. \and Troyansky L.}
  \REVIEW{Nature}{400}{1999}{133}.

\bibitem{Weinberger1993}
\Name{Weinberger E.~D. \and Stadler P.~F.} \REVIEW{J. Theor.
  Biol.}{163}{1993}{255}.

\bibitem{Gross1984}
\Name{Gross D.~J. \and M\'ezard M.} \REVIEW{Nucl. Phys. B}{240}{1984}{431}.

\bibitem{Gardner1985}
\Name{Gardner E.} \REVIEW{Nucl. Phys. B}{257}{1985}{747}.

\bibitem{CrSo1992}
\Name{Crisanti A. \and Sommers H.-J.} \REVIEW{Z. Phys. B: Cond.
  Matt.}{87}{1992}{341}.

\bibitem{CrSo1995}
\Name{Crisanti A. \and Sommers H.-J.} \REVIEW{J. Phys. I France}{5}{1995}{805}.

\bibitem{PhysRevB.57.11251}
\Name{Cavagna A., Giardina I. \and Parisi G.} \REVIEW{Phys. Rev.
  B}{57}{1998}{11251}.

\bibitem{PhysRevLett.71.173}
\Name{Cugliandolo L.~F. \and Kurchan J.} \REVIEW{Phys. Rev.
  Lett.}{71}{1993}{173}.

\bibitem{Stariolo1990}
\Name{Stariolo D.~A.} \REVIEW{Physica A}{166}{1990}{622}.

\bibitem{Talagrand2003}
\Name{Talagrand M.} \REVIEW{Rev. Math. Phys.}{15}{2003}{1}.

\bibitem{Nakajima2008}
\Name{Nakajima T. \and Hukushima K.} \REVIEW{J. Phys. Soc.
  Japan}{77}{2008}{074718}.

\bibitem{Agliari2012a}
\Name{Agliari E., Barra A., Burioni R. \and {Di Biasio} A.} \REVIEW{J. Math.
  Phys.}{53}{2012}{063304}.

\bibitem{Janis2015}
\Name{Jani{\v{s}} V., Kauch A. \and Kl{\'{i}}{\v{c}} A.} \REVIEW{Phase
  Trans.}{88}{2015}{245}.

\bibitem{Rieger1992}
\Name{Rieger H.} \REVIEW{Phys. Rev. B}{71}{1992}{14655}.

\bibitem{Oliveira1997}
\Name{de~Oliveira V.~M. \and Fontanari J.~F.} \REVIEW{J. Phys.
  A}{30}{1997}{8445}.

\bibitem{Montanari2003}
\Name{Montanari A. \and Ricci-Tersenghi F.} \REVIEW{Eur. Phys. J.
  B}{33}{2003}{339}.

\bibitem{Crisanti2005a}
\Name{Crisanti A., Leuzzi L. \and Rizzo T.} \REVIEW{Phys. Rev.
  B}{71}{2005}{094202}.

\bibitem{Rizzo2013a}
\Name{Rizzo T.} \REVIEW{Phys. Rev. E}{88}{2013}{032135}.

\bibitem{Monasson1995}
\Name{Monasson R.} \REVIEW{Phys. Rev. Lett.}{75}{1995}{2847}.

\bibitem{Barrat1996b}
\Name{Barrat a., Burioni R. \and M{\'{e}}zard M.} \REVIEW{J. Phys.
  A}{29}{1996}{L81}.

\bibitem{Franz1995a}
\Name{Franz S. \and Parisi G.} \REVIEW{J. Phys. I (France)}{5}{1995}{21}.

\bibitem{KiTh1987a}
\Name{Kirkpatrick T.~R. \and Thirumalai D.} \REVIEW{Phys. Rev.
  Lett.}{58}{1987}{2091}.

\bibitem{Ferrari2012}
\Name{Ferrari U., Leuzzi L., Parisi G. \and Rizzo T.} \REVIEW{Phys. Rev.
  B}{86}{2012}{014204}.

\bibitem{Crisanti2000}
\Name{Crisanti A. \and Ritort F.} \REVIEW{Eurphys. Lett.}{51}{2000}{147}.

\bibitem{Crisanti2000a}
\Name{Crisanti A. \and Ritort F.} \REVIEW{Europhys. Lett.}{52}{2000}{640}.

\bibitem{Bouchaud1992}
\Name{Bouchaud J.-P.} \REVIEW{J. Phys. I (France)}{2}{1992}{1705}.

\bibitem{Arous2002}
\Name{{Ben Arous} G., Bovier A. \and Gayrard V.} \REVIEW{Phys. Rev.
  Lett.}{88}{2002}{087201}.

\bibitem{Arous2008}
\Name{{Ben Arous} G., Bovier A. \and {\v{C}}ern{\'{y}} J.} \REVIEW{J. Stat.
  Mech.}{2008}{2008}{L04003}.

\bibitem{Gay2016}
\Name{Gayrard V.} \Book{Aging in metropolis dynamics of the rem: a proof}
  arXiv:1602.06081 (2016).

\bibitem{Cerny2017}
\Name{{\v{C}}ern{\'{y}} J. \and Wassmer T.} \REVIEW{Probability Theory and
  Related Fields}{167}{2017}{253:303}.

\bibitem{Dyre1987}
\Name{Dyre J.~C.} \REVIEW{Phys. Rev. Lett.}{58}{1987}{792}.

\bibitem{BouchaudDean1995}
\Name{Bouchaud J.-P. \and Dean D.~S.} \REVIEW{J. Phys. I
  (France)}{5}{1995}{265}.

\bibitem{Monthus_1996}
\Name{Monthus C. \and Bouchaud J.-P.} \REVIEW{J. Phys. A: Math.
  Gen.}{29}{1996}{3847}.

\bibitem{Cammarota2018}
\Name{Cammarota C. \and Marinari E.} \REVIEW{J. Stat.
  Mech.}{2018}{2018}{043303}.

\bibitem{BenArous2003}
\Name{{Ben Arous} G., Bovier A. \and Gayrard V.} \REVIEW{Comm. Math.
  Phys.}{235}{2003}{379}.

\bibitem{Bovier2003}
\Name{{Ben Arous} G., Bovier A. \and Gayrard V.} \REVIEW{Comm. Math.
  Phys.}{236}{2003}{1}.

\bibitem{Arous2008a}
\Name{{Ben Arous} G., Bovier A. \and {\v{C}}ern{\'{y}} J.} \REVIEW{Comm. Math.
  Phys.}{282}{2008}{663}.

\bibitem{Rinn2000}
\Name{Rinn B., Maass P. \and Bouchaud J.-P.} \REVIEW{Phys. Rev.
  Lett.}{84}{2000}{5403}.

\bibitem{Doliwa2005}
\Name{Heuer A., Doliwa B. \and Saksaengwijit A.} \REVIEW{Phys. Rev.
  E}{72}{2005}{021503}.

\bibitem{Baity-Jesi2018}
\Name{Baity-Jesi M., Biroli G. \and Cammarota C.} \REVIEW{J. Stat.
  Mech.}{2018}{2018}{013301}.

\bibitem{Baity-Jesi2018a}
\Name{Baity-Jesi M., {Achard-de Lustrac} A. \and Biroli G.} \REVIEW{Phys. Rev.
  E}{98}{2018}{012133}.

\bibitem{Billoire2005}
\Name{Billoire A., Giomi L. \and Marinari E.} \REVIEW{Europhys.
  Lett.}{71}{2005}{824}.

\bibitem{Cugliandolo1997}
\Name{Cugliandolo L.~F., Kurchan J., {Le Doussal} P. \and Peliti L.}
  \REVIEW{Phys. Rev. Lett.}{78}{1997}{350}.

\bibitem{Iguain2001}
\Name{Berthier L., Cugliandolo L.~F. \and Iguain J.~L.} \REVIEW{Phys. Rev.
  E}{63}{2001}{051302}.

\bibitem{Berthier2003}
\Name{Berthier L.} \REVIEW{J. Phys.: Condens. Matter}{15}{2003}{S933}.

\bibitem{Bolthausen2014}
\Name{Bolthausen E.} \REVIEW{Comm. Math. Phys.}{325}{2014}{333}.

\bibitem{Bouchaud1997}
\Name{Bouchaud J.-P. \and M{\'{e}}zard M.} \REVIEW{J. Phys. A}{30}{1997}{7997}.

\bibitem{Carpentier2001}
\Name{Carpentier D. \and Le~Doussal P.} \REVIEW{Phys. Rev.
  E}{63}{2001}{026110}.

\bibitem{CluselBertin2008}
\Name{Clusel M. \and Bertin E.} \REVIEW{Int. Jour. Mod. Phys.
  B}{22}{2008}{3311}.

\end{thebibliography}

\end{document}